\documentclass[12pt]{article}
\newcommand{\beq}{\begin{equation}}
\newcommand{\eeq}{\end{equation}}
\newcommand{\bea}{\begin{eqnarray}}
\newcommand{\eea}{\end{eqnarray}}
\newcommand{\du}{d_{\cal U}}
\newcommand{\dbz}{d_{BZ}}
\newcommand{\half}{\frac{1}{2}}

\newcommand{\lagr}{{\cal L}}
\newcommand{\opr}{{\cal O}}
\newcommand{\unp}{{\cal U}}
\newcommand{\Lu}{\Lambda_\unp}
\newcommand{\bra}{\langle}
\newcommand{\ket}{\rangle}
\newcommand{\Mbh}{M_{\rm BH}}
\newcommand{\Mpl}{M_{\rm Pl}}
\newcommand{\Msol}{M_{\odot}}
\newcommand{\Mgut}{M_{\rm GUT}}
\newcommand{\Mpbh}{M_{\rm PBH}}

\usepackage{graphicx}
 \usepackage{setspace}
 \usepackage{amsfonts}
 \usepackage{amsmath}  
\begin{document}
\pagestyle{empty}
\begin{center}
\vskip 2cm
{\bf \LARGE Constraints on Vector Unparticle Physics from Cosmic Censorship}
\vskip 1cm
J. R. Mureika \\

{\small \it Department of Physics, Loyola Marymount University, Los Angeles, CA  90045-2659} \\
Email: jmureika@lmu.edu
\end{center}

{\noindent{\bf Abstract} \\
Vector unparticle couplings to standard model fields produce repulsive corrections to gravity.  From a general relativistic perspective, this leads to an effective Reissner-Nordstr\"om-like metric whose ``charge'' is a function of the unparticle coupling constant $\lambda$, and therefore can admit naked singularities.   Requiring the system to respect cosmic censorship  provides a new method of constraining the value of $\lambda$.  These limits are extremely loose for stellar-mass black holes, but commensurate with existing bounds for primordial black holes.  In the case of theoretical low-mass black holes, the bounds on $\lambda$ are much stricter than those derived from astrophysical and accelerator phenomenology.    Additional constraints on the lower limit of $\lambda$ are used to estimate the mass of the smallest possible black hole $\Mbh^{\rm min}$ that can be formed in the unparticle framework, as a function of the unparticle parameters ($\Lambda_\unp,M_\unp,\du,\dbz$).
}


\section{Introduction}
The unparticle ``revolution'' \cite{georgi} -- a new paradigm for physics beyond the standard model (SM) --  has prompted a wealth of speculative literature in both particle phenomenology and cosmology.   Its framework consists of a weakly-coupled Banks-Zaks (BZ) field \cite{bz} that exchanges with the SM a massive particle $M_{\unp}$  via suppressed non-renormalizable interactions 
\beq
\lagr = \frac{1}{M^k_{\unp}} \opr_{SM}\opr_{BZ}~~.
\eeq
The field operators $\opr_{SM},\opr{BZ}$ have dimensions $d_{SM}$ and $d_{BZ}$ respectively, which necessitates $k = d_{SM} + d_{BZ} - 4$. 

Below some energy scale $\Lambda_\unp < M_\unp$, the coupling begins to run as the BZ field undergoes dimensional transmutation to become ``unparticle stuff'', represented by the operator $\opr_\unp$ of dimension $d_\unp \ne d_{BZ}$.  This leads to a new interaction picture represented by the Lagrangian term
\beq
\frac{\Lambda_\unp^{\dbz-\du}}{M_\unp^{\dbz +d_{SM}-4}} \opr_\unp \; \opr_{SM}~~.
\eeq 
Postulating that $\Lambda_\unp \sim 1~$TeV provides a wealth of new physics that can not only be observabled at the LHC, but also stands to modify astrophysical and cosmological mechanisms.

The matrix element 
\beq
\bra 0 | \opr_\unp(x) \opr^\dagger_\unp(0)|0\ket  = \int \frac{d^4p}{(4\pi)^4} e^{iPx} |\bra 0 | \opr_\unp(0)|P\ket|^2 \rho(P^2)
\eeq
for an unparticle of four-momentum $P$ constrains the spectral density function to be of the form
\beq
|\bra 0 | \opr_\unp(0)|P\ket|^2 \rho(P^2) = {A_d}_{\cal U} \theta(P^0) \theta(P^2) (P^2)^{\du -2}~,
\eeq
where
\beq
A_{\du} = \frac{16 \pi^{5/2}}{(2\pi)^{2n}} \frac{\Gamma(n+1/2)}{\Gamma(n-1)\Gamma(2n)}~~.
\eeq
When compared to the standard phase space $A_n \theta(P^0) \theta(P^2) (P^2)^{n -2}$ of $n$ interacting particles of total momentum $P$, it can be concluded that unparticles behave as a system of $\du$ (non-integer) fundamental particle states \cite{georgi}.  

Several explanations of the physical nature of unparticle stuff include a composite Banks-Zaks particle with a continuum of masses \cite{kraz,nikolic,mcdonald2}.  Such an interpretation is indeed {\it un}particle in nature: it defies the notion of discrete fundamental mass eigenstates that rests at the heart of modern quantum theory.  A more tradition field-theoretic approach has been proposed in \cite{georgi2,georgi3}, whereby the unparticle phase space is constructed from a Sommerfield model of massless fermions coupled to a massive vector field.  

Unparticles may possess any of the standard Lorentz signatures (scalar, vector, tensor, or spinor).  This paper will focus specifically on the corrections to gravitation supplied by vector unparticle matter.  It will be shown that fundamental limits may be placed on both the unparticle parameter space, as well as black hole physics in general.

\section{Vector Unparticle Physics and Ungravity}
\begin{figure}[h]
\begin{center}
\leavevmode
\includegraphics[scale=0.6]{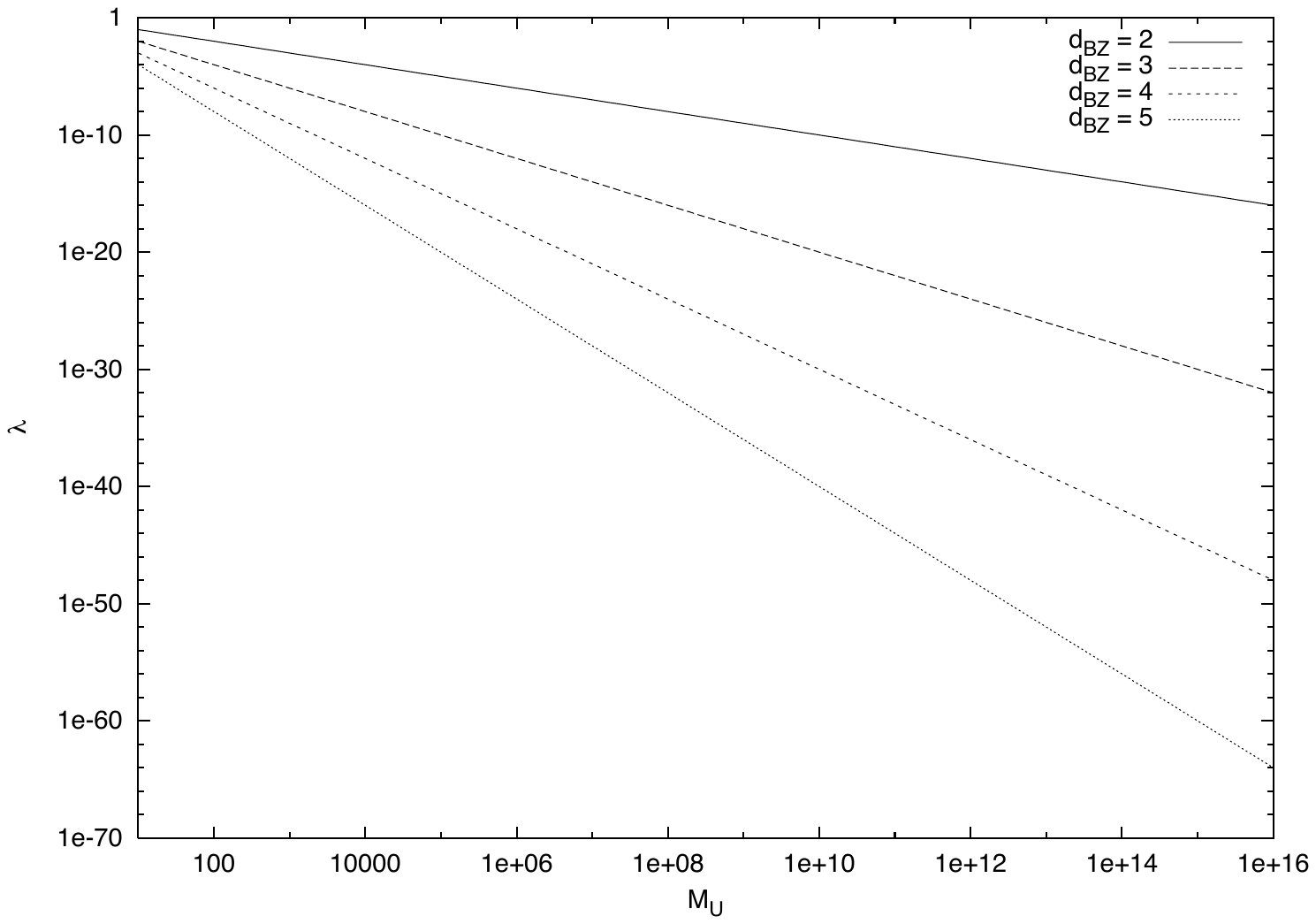}
\caption{Magnitude of the vector unparticle coupling constant $\lambda$ as a function of energy scale $M_\unp$ and operator dimension $\dbz$, with $\Lambda_\unp = 1~$TeV.  Higher values of $\Lambda_\unp$ will increase the lower bound of the coupling.  The minimum value of $\lambda$ is defined by Equation~\ref{lambdadef}.}
\label{fig0}
\end{center}
\end{figure}

Vector unparticles couple to baryon currents $J_\mu$ via the interaction
\beq
\lagr \sim \frac{\lambda}{\Lambda_\unp^{\du-1}} J_\mu \opr^\mu_\unp ~~,
\label{veccurrent}
\eeq
where $\lambda$ is the dimensionless coupling constant \cite{hsu1,damora}
\beq
\lambda \sim C \left(\frac{\Lambda_\unp}{M_\unp}\right)^{\dbz - 1}
\label{lambdadef}
\eeq
and $C$ is a parameter of order unity.  The energy scale hierarchy $\Mpl \geq M_\unp > \Lambda_\unp \geq 1~$TeV necessitates $\lambda < 1$, except in the degenerate case when $\Lambda_\unp \rightarrow M_\unp$.  Since it is generally assumed $M_\unp \leq \Mpl$, the relative size of $\Lambda_\unp$ and the Banks-Zaks dimension $\dbz$ set the possible lower bounds on $\lambda$.   Figure~\ref{fig0} demonstrates the possible range of the coupling strength.

Conversely, upper bounds on its magnitude are derived from phenomenological considerations.  The literature suggests a wide range of such bounds, from as large as $\lambda < 1$ to as small as $\lambda < 10^{-20}$.  A limited, but certainly not exhaustive, list of examples include constraints from solar system physics \cite{damora}, collider data \cite{cheung1,cheung2}, neutrino phenomenology \cite{garcia}, and big bang nucleosynthesis \cite{bertolami}.   A large hierarchy between $\Lambda_\unp$ and $M_\unp$ allows for smaller values of $\dbz$, whereas a smaller hierarchy necessitates larger values of $\dbz$ to achieve the bounds on $\lambda$ cited in the literature.  Note that many studies consider only the extreme case $\lambda = 1$ \cite{cheung1}.  Non-commutative geometry \cite{nicolini1,nicolini2} has been suggested as an additional manifestation of unparticle physics, from which parameter bounds may be extracted.

The modifications of classical gravitational laws via interactions of SM particles with unparticle stuff can be readily calculated.  Pioneering approaches and applications of this phenomenology can be found in \cite{damora,goldberg,hsu1,liao,jrm1} and references therein.     It was originally shown that, if treated as a perturbative extension of gravitation, unparticle physics may modify the effective metric structure of spacetime \cite{jrm1,jrm2}.    In the case of scalar and tensor unparticles, this solution can be written
\bea
ds^2 = \left[1-\frac{2GM}{r}\left(1+ \left(\frac{R_{s,t}}{r}\right)^{2\du-2}\right)\right] \; dt^2 -
 \frac{dr^2}{1-\frac{2GM}{r}\left(1+\left(\frac{R_{s,t}}{r}\right)^{2\du-2}\right)} - r^2 d\Omega^2 ~~,
\label{unpmetric}
\eea
with the fundamental scalar (tensor) length scale $R_{s,t}$ set by the unparticle energies $\Lambda_\mu, M_\mu$ and dimensions $\du,\dbz$.  For SM interaction distances $r \ll R_{s,t}$, the unparticle coupling mimics extra-dimensional physics with a similar compactification radius.   In particular, this can lead to the formation of microscopic black holes in high energy collisions.  

It has since been shown \cite{euro2} the metrics (\ref{unpmetric}) can be derived as an exact solution to the Einstein equations from a non-local action of the form \cite{euro1}
\beq
S_\unp = -\int~d^4x~\sqrt{g} \left[1 + \frac{A_{\du}}{(2\du-1)\sin(\pi \du)} \frac{\kappa_*^2}{\kappa^2} \left(\frac{-D^2}{\Lu^2}\right)^{1-\du}\right]^{-1} R~~,
\label{euroaction}
\eeq
with $D^2$ the generally covariant derivative operator, $R$ the Ricci scalar, and $\kappa_* = \Lu^{-1} (\Lu/M_\unp)^{\dbz}$ the ungravity coupling constant.  The resulting equations of motion manifest the unparticle theory, which have been cast in the scalar \cite{euro2} and vector forms \cite{euro3}.  Such non-local actions are of increasing interest in the literature.  A similar action of this form has been shown to provide ultraviolet completeness of a regular and noncommutative quantum gravity theory \cite{euro4,nicolini3}, and the black hole solutions from a higher-dimensional version (\ref{euroaction}) are currently being explored \cite{jrm3}.

Variation of (\ref{euroaction}) leads to the Einstein equations \cite{euro2}
\beq
R^\mu_\nu - \frac{1}{2}\delta^\mu_\mu = \kappa^2 \left[1+ \frac{A_{\du} \Lu^{2-2\du}}{(2\du-1)\sin(\pi \du)} \frac{\kappa^2_*}{\kappa^2}(-D)^{\du-1}\right] T^\mu_\nu
\eeq
whose stress-energy tensor has the form 
\beq
T^0_0 = -\frac{M}{4\pi r^2} \delta(r)~~,~~T^r_r = 0~~,~~T^\theta_\theta = T^\phi_\phi = -\frac{M}{16\pi r}\delta(r) \frac{1}{g_{00}}\partial_r g_{00}
\label{tmunu}
\eeq
The $(00)-$ and $(rr)-$metric components can be shown to have the expected reciprocal symmetry \cite{euro2},
\beq
g_{rr} = \left(1-\frac{2GM(r)}{r}\right)^{-1} \propto -\frac{1}{g_{00}}
\eeq
with $M(r)$ a radial function related to the mass of the source.   For regular gravity, this reduces to the source mass $M_{BH}$ and the pure Schwarzschild limit.  When constrained by the stress-energy (\ref{tmunu}), the function $M(r)$ exactly reproduces the unparticle black hole solutions discussed in \cite{jrm1,jrm2}.

%In the case of scalar and tensor unparticles, this solution can be written
%\bea
%ds^2 = \left[1-\frac{2GM}{r}\left(1+ \left(\frac{R_{s,t}}{r}\right)^{2\du-2}\right)\right] \; dt^2 -
% \frac{dr^2}{1-\frac{2GM}{r}\left(1+\left(\frac{R_{s,t}}{r}\right)^{2\du-2}\right)} - r^2 d\Omega^2 ~~,
%\label{unpmetric}
%\eea
%with the fundamental scalar (tensor) length scale $R_{s,t}$ set by the unparticle energies $\Lambda_\mu, M_\mu$ and dimensions $\du,\dbz$.  For SM interaction distances $r \ll R_{s,t}$, the unparticle coupling mimics extra-dimensional physics with a similar compactification radius.   In particular, this can lead to the formation of microscopic black holes in high energy collisions.  

For vector unparticle interactions, a repulsive correction to the standard Newtonian gravitational potential is introduced \cite{damora,hsu1},
\beq
V(r) = -\frac{GmM}{r} \left[ 1 - \left(\frac{R_v}{r}\right)^{2\du-2}\right]~~
\eeq
where the interaction length $R_v$ is
\beq
R_v = \left[\frac{1}{2\pi^{2\du}} \; \frac{\Gamma(\du+\half)\Gamma(\du-\half)}{\Gamma(2\du)}\right]^\frac{1}{2\du-2} \left(\frac{\lambda \Mpl}{u}\right)^\frac{1}{\du-1} \Lambda_\unp^{-1}
\label{vecscale}
\eeq
Here.  $u \sim 1~$GeV is the baryon mass, introduced via the coupling to the current in Equation~\ref{veccurrent}. The metric then assumes the form \cite{euro3}
\beq
g_{00} =  1-\frac{2G\Mbh}{r}\left(1- \left(\frac{R_v}{r}\right)^{2\du-2}\right)~~,~~g_{11} = -g_{00}^{-1}
\label{vecmetric}
\eeq
While the repulsive nature of the correction does not strengthen gravitational attractions, it nevertheless introduces additional black hole-related novelties.   Specifically, multiple horizons can be obtained without the need for electric charge or angular momentum.

This result is valid for arbitrary non-perturbative situations and may be formally derived from the action (\ref{euroaction}) by an appropriate modification of the coupling constant $\kappa_*^2$ for vector unparticle interactions.  This amounts to a Lorentz signature-dependent sign change of the form  $\kappa_*^2 \rightarrow (-1)^s \kappa_*^2$, , where $s= \{0,1,2\}$ for $\{$scalar, vector, tensor$\}$ unparticles respectively \cite{euro3}. 

It should be noted the metrics derived from this non-local action (9) are often criticized for their failure to respect Birkhoff's theorem, in that the corresponding solutions do not represent a true vacuum (and thus the uniqueness of obvious black hole solutions could be doubted).  In
reality, this rigid constraint is misplaced in the current context, as it is predicated on the stipulation that the metric stems from the Einstein-Hilbert action.  
Since the unparticle-enhanced action represents a different (and heretofore unexplored) spacetime, there is no reason to expect such a uniqueness theorem to be
applicable.

%Any non-zero curvature sourced by some arbitrary mass distribution -- including the classic Schwarzschild and other black hole solutions -- should logically be represented by an action of the form (\ref{euroaction}).  True vacuum solutions are intrisically flat, since by definition there is nothing to induce curvature \cite{euro2}.  This amounts to a Lorentz signature-dependent sign change of the form  $\kappa_*^2 \rightarrow (-1)^s \kappa_*^2$, , where $s= \{0,1,2\}$ for $\{$scalar, vector, tensor$\}$ unparticles respectively \cite{euro3}. 

\section{Vector Ungravity-Enhanced Black Holes and Cosmic Censorship}
\subsection{The Reissner-Nordstr\"om case}
As a simple example, consider the case $\du = 1.5$.  The metric coefficients are
\beq
g_{00} =  1-\frac{2G\Mbh}{r} + \frac{2G\Mbh R_v}{r^2}~~
\label{vecmetric2}
\eeq
which is immediately recognizable as a Reissner-Nordstr\"{o}m-like metric whose ``charge'' is determined by the vector unparticle scale $R_v$.    It can be shown that the Ricci scalar vanishes in this limit, providing a true vacuum solution that respects Birkhoff's theorem.  This admits two non-zero horizon solutions
\beq
r = G\Mbh \pm \sqrt{G^2\Mbh^2-2G\Mbh R_v}~~
\label{rnsol}
\eeq
whenever $G\Mbh > 2R_v$.  In the limit $G\Mbh \gg 2R_v$, the solution approaches the Schwarzschild radius, and when $G\Mbh = 2R_v$ the black hole is extremal.

If $G\Mbh < 2R_v$, the solution (\ref{rnsol}) becomes complex and yields no horizons.  The remaining singularity at $r=0$ is naked.    An appeal to the cosmic censorship hypothesis \cite{penrose} negates the possible existence of such defects (their formation is forbidden by considering {\it e.g.} linearized perturbations of the Schwarzschild metric \cite{wald}).  

From this perspective, the condition $2R_v < G\Mbh$ can be understood to be a fundamental constraint.  Since the interaction scale $R_v$ is explicitly a function of the vector unparticle parameters (Equation~\ref{vecscale}), this condition provides a new method of gauging their values, and therefore restricting the size of the coupling strength $\lambda$.  

Numerically, Equation~\ref{lambdadef} can be written
\beq
R_v =1.38\times 10^{36} \; \lambda^2 \; \Lambda_\unp^{-1}~~,
\eeq
and the censorship condition becomes
\beq
\lambda \leq 6\times 10^{-19} \sqrt{\frac{\Lambda_\unp \Mbh}{\Mpl^2}}
\eeq
where the replacement $G = \Mpl^{-2}$ has been made.  Using the minimum value of $\Lambda_\unp = 1~$TeV, this constraint is
\beq
\lambda <6\times 10^{-35} \sqrt{\frac{\Mbh}{1~\rm TeV}}~~.
\label{lapprox}
\eeq    

To date, indirect observational evidence has been found for roughly 20 black holes in the range $\Mbh \sim M_\odot =10^{30}~$kg \cite{kovacs}, including the pioneering Cyg-X1 \cite{bolton} ($\Mbh > 4\Msol$), XTE J1650-500 \cite{orosz} ($\Mbh \sim 4\Msol$), and GRO J0422+32 \cite{gelino} ($\Mbh \sim 3-5 \Msol$). For such stellar mass-sized black holes, one finds $\lambda < 7\times 10^{-8}$.    According to Equation~\ref{lapprox}, the upper bound on $\lambda$ depends on $\sqrt{\Lambda_\unp}$.  Thus, increasing the possible value of $\Lambda_\unp$ will loosen these constraints.  Indeed, if $\Lambda_\unp = \Mgut \sim 10^{13}~$TeV, then $\lambda < 0.22$ for $\Mbh=M_\odot$.

Further constraints can be derived from primordial black holes (PBH) of mass $\Mbh \sim 10^{12}~$kg that are believed to have formed following the inflationary era.  These have since evaporated but can be ``detected''  through their signatures in the stochastic gravitational wave background \cite{pbh1,pbh2,pbh3}.   At this mass threshold, the limits on the coupling are $\lambda < 5\times 10^{-17}$ for $\Lambda_\unp = 1~$TeV, and $\lambda < 1.5\times 10^{-10}$ when $\Lambda_\unp = \Mgut$.   These limits are in excellent agreement with those in the current literature, derived from a variety of mechanisms. 

\subsection{Preservation of unitarity}
The above example, while useful as an simple illustration of this research, does not consider unitarity constraints imposed by unparticle physics.  Only scalar unparticles processes may occur with dimensions $\du \geq 1$.
In fact, to avoid such unitarity violations the scaling dimension for vector unparticles must be $\du \geq 3$ \cite{yu,grinstein}. 

For arbitrary dimension $\du$, the black hole horizons are roots of the equation
\beq
0= r^{2\du-1} -\frac{2\Mbh}{\Mpl^2}\left(r^{2\du-2}-R_v^{2\du-2}\right)~~.
\label{horizons}
\eeq
Although there are no analytical solutions in this case, limits on the value of $R_v$ (and hence $\lambda$) may easily be obtained numerically.   In all cases, two real positive solutions $r_\pm$ can be obtained by varying $\lambda$ and/or $\Mbh$, whose behavior is in accordance with that expected for the Reissner-Nordstr\"om case.  As expected, the solutions converge to a common horizon where the black hole solution is  extremal, beyond which all solutions are complex and the singularity is naked.

\begin{figure}[h]
\begin{center}
\leavevmode
\includegraphics[scale=0.6]{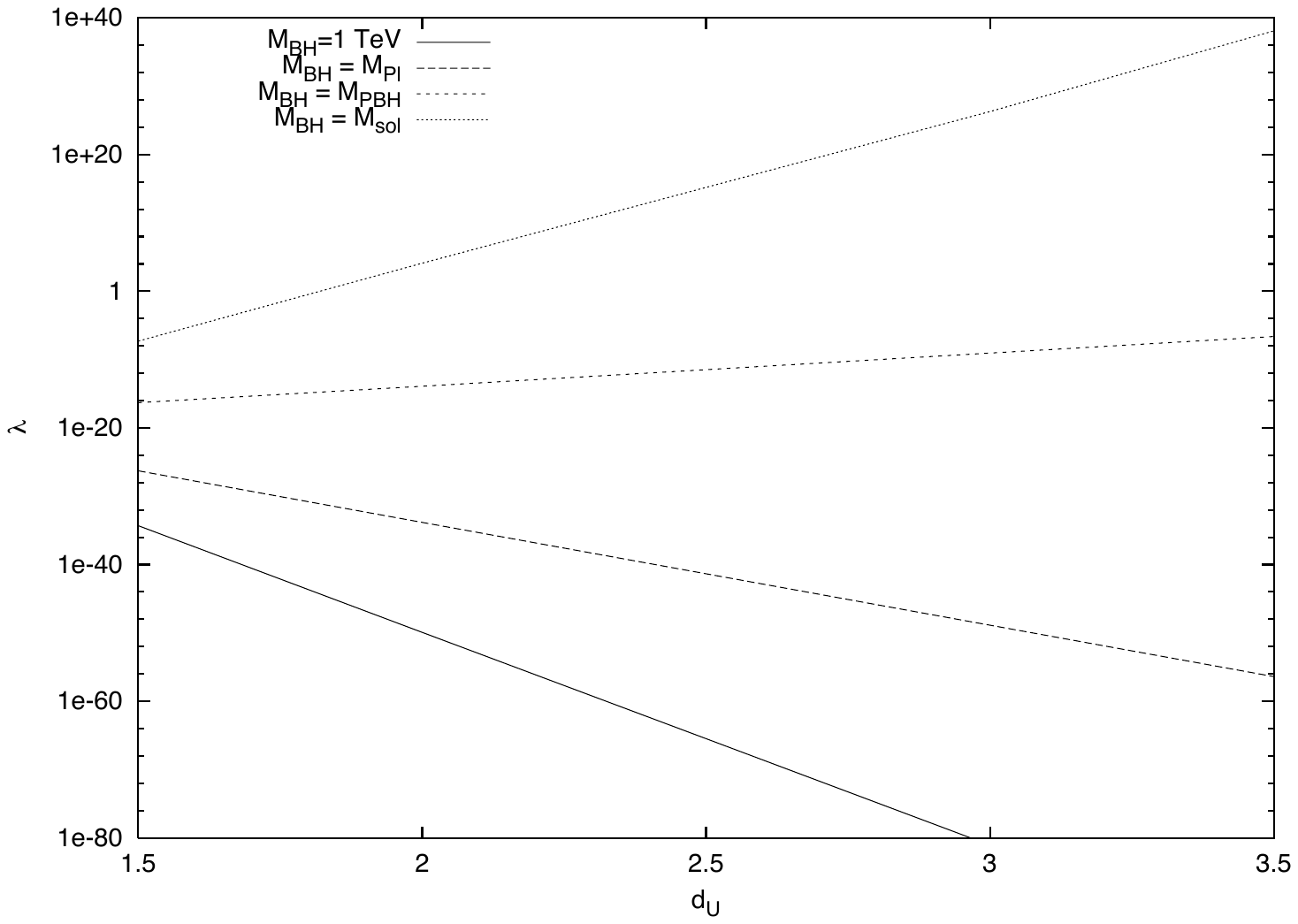}
\caption{Allowed unparticle parameter space $(\du, \lambda)$ for $\Lambda_\unp = 1~$TeV as a function of black hole mass $\Mbh$ ($M_{\rm PBH} = 10^{12}~$kg, $\Msol = 10^{30}~$kg).  Allowed regions are bounded from above by $\lambda = 1$, and by the hierarchy constraint in Equation~\ref{lambdadef}.}
\label{fig1}
\end{center}
\end{figure}

Figure~\ref{fig1} shows the critical values of $\lambda$ above which cosmic censorship is violated.   This condition can potentially narrow the allowed parameter space from that suggested by previous investigations.  Constraints from solar-mass black holes are so outrageously large ($\lambda \gg 1$) as to be considered useless in the face of existing limits.  While this might be perceived an unfortunate failure of the approach, in a sense it is akin to putting limits on the fundamental unit of electric charge from measurements of astrophysical black holes.  

Primordial black holes ($\Mpbh \sim 10^{12}~$kg), on the other hand, satisfy $\lambda < 1$ for a suitable range of $\du \geq 3$.   When $\du = 3$, one finds $\lambda < 9\times 10^{-10}$, and $\lambda < 2\times 10^{-7}$ for $\du = 3.5$ (where $\Lambda = 1~$TeV in each case).  Again, these values are generally commensurate with the largest limits on $\lambda$ available in the literature.  Observation evidence of PBHs, then, provides a useful and novel contribution to the unparticle parameter space limits.  As the scale $\Lambda_\unp$ increases, however, so does the upper bound of $\lambda$ (and eventually surpasses the saturation bound $\lambda =1$).

\subsection{The littlest black hole}
\begin{figure}[h]
\begin{center}
\leavevmode
\includegraphics[scale=0.6]{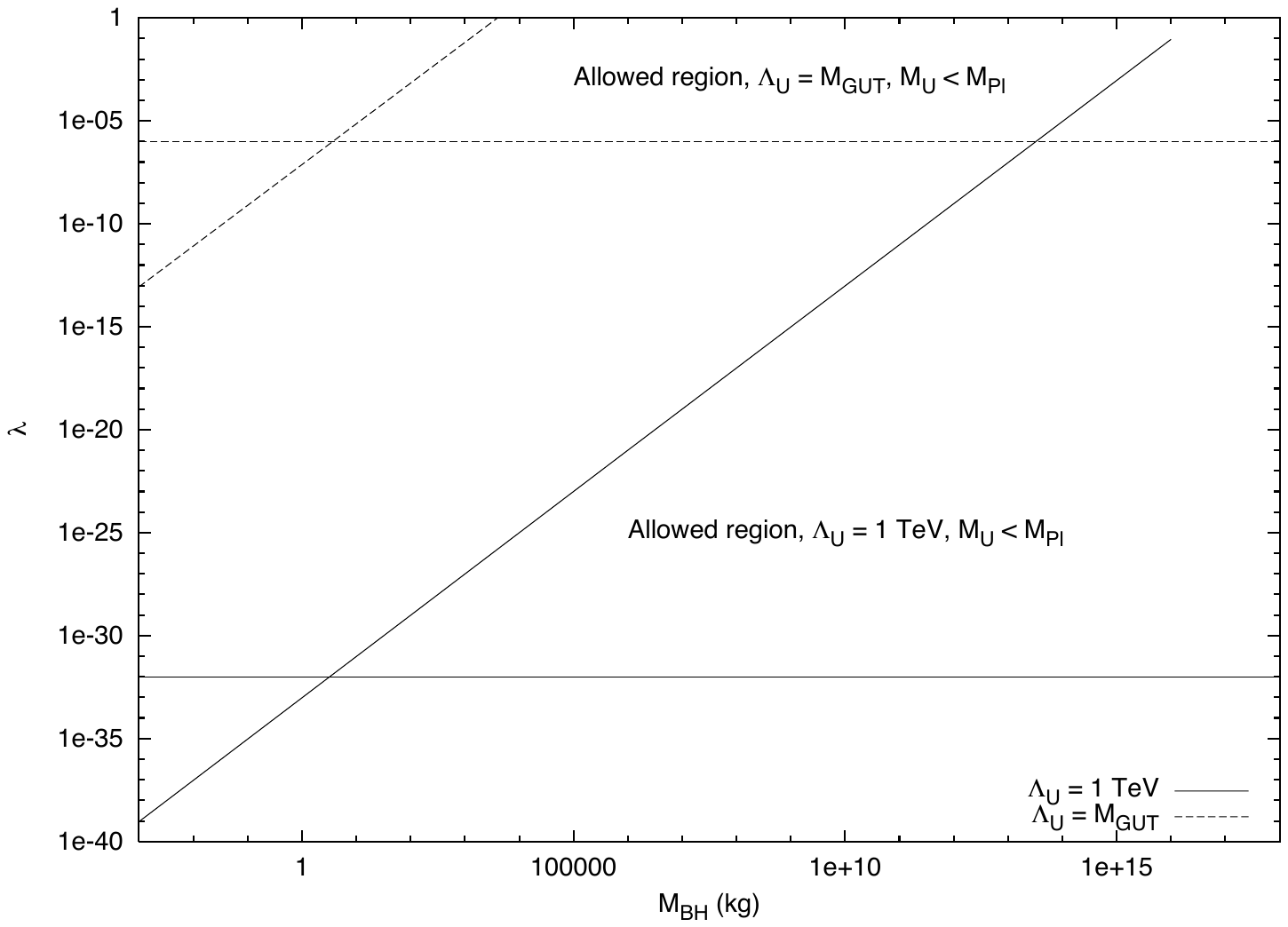}
\caption{Limits on minimum (extremal) black hole mass $\Mbh$ for $\du = 3,\dbz=3$, with ranges on the unparticle energy scale $\Lambda_\unp  = 1~{\rm TeV} - 10^{13}~$TeV ($M_{\rm GUT}$).   The parameter space is always bounded from above by $\lambda=1$, in which case $\Lambda_\unp = M_\unp = \Mpl$.}
\label{fig2}
\end{center}
\end{figure}

\begin{figure}[h]
\begin{center}
\leavevmode
\includegraphics[scale=0.6]{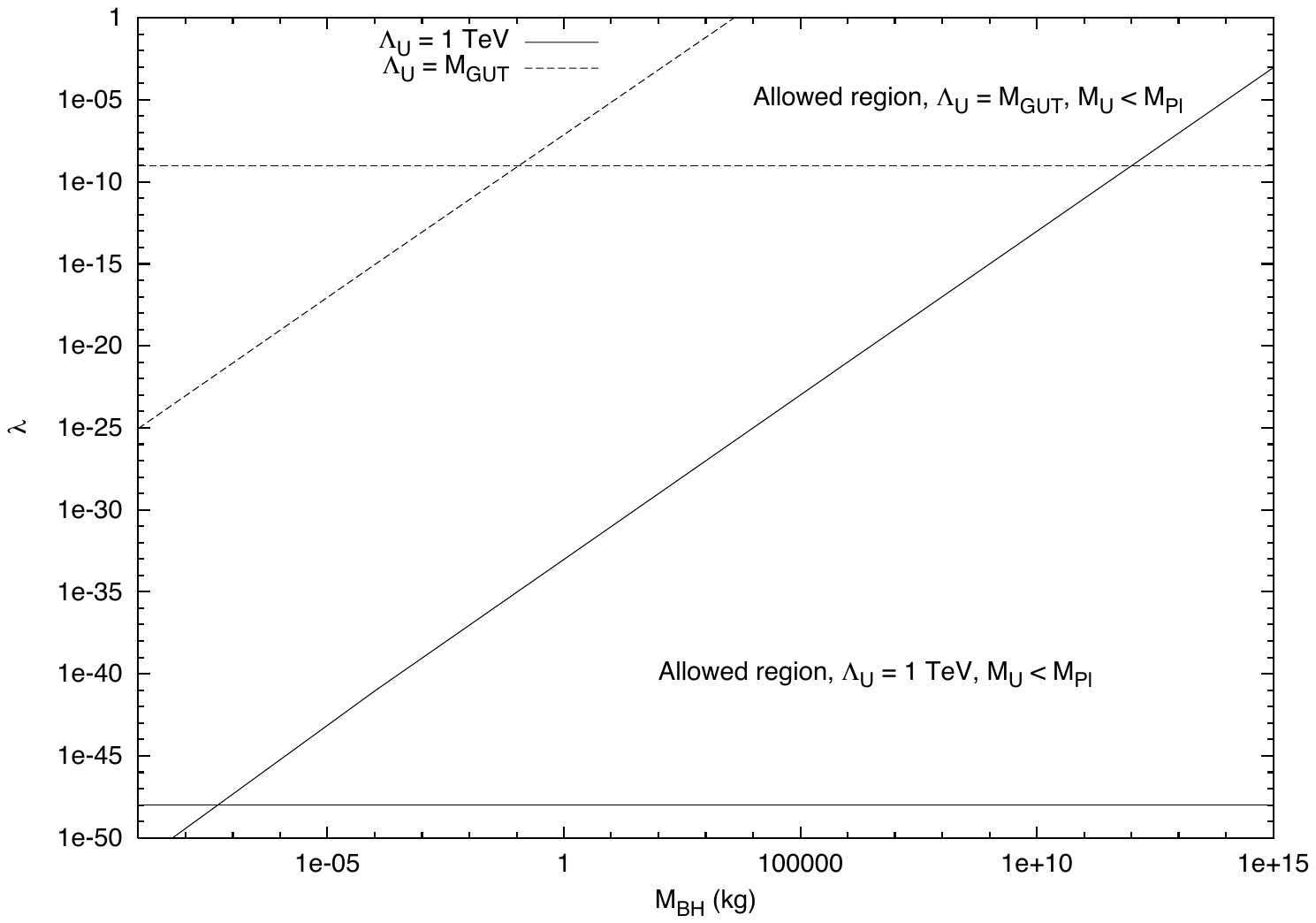}
\caption{Limits on minimum (extremal) black hole mass $\Mbh$ for $\du = 3, \dbz=4$, with ranges on the unparticle energy scale $\Lambda_\unp  = 1~{\rm TeV} - 10^{13}~$TeV ($M_{\rm GUT}$).}
\label{fig3}
\end{center}
\end{figure}

From a more conjectural approach, one can instead focus on the constraints provided to ultra-low-mass (even microscopic) black holes.    That is, the limits on $\lambda$ can be used to determine the minimum mass of a black hole that can form in this framework.  Figures~\ref{fig2} and ~\ref{fig3} shows how maximum value of 
$\lambda$ varies with black hole mass $\Mbh$, for the possible range of $\Lambda_\unp$ values, subject to variation in the Banks-Zaks dimension $\dbz$.  

The value of $M_\unp$ is calculated from Equation~\ref{lambdadef}.   Increasing values of $\Mbh$ correspond to increased upper limits on $\lambda$.  The value $\lambda =1$ is the greatest upper bound of the parameter space, since this represents the degenerate case $\Lambda_\unp = M_\unp$.  As $\Lambda_\unp$ increases in value, the ratio that defines $\lambda$ (Equation~\ref{lambdadef}) also approaches 1, since $\Lambda_\unp < M_\unp \leq \Mpl$,  Much higher values of $\dbz$ are required to produce smaller values of $\lambda$ (see Figure~\ref{fig1}), and thus the physical reality of such a scenario seems less likely.     

Since the value of $\lambda$ is constrained by Equation~\ref{lambdadef}, however, the parameter space may be further narrowed by fixing the value of $\dbz$ (see Figures~\ref{fig2} and~\ref{fig3}).  If $\dbz = 3$ and $\Lambda_\unp = 1~$TeV, then $\lambda \sim \left(\frac{1~\rm TeV}{M_\unp}\right)^2$.  Assuming that $M_\unp$ can be no larger than $\Mpl$, the minimum possible value of the coupling is $\lambda_{\rm min} \sim 10^{-32}$.  For $\dbz = 4$, the lower bound extends to $\lambda_{\rm min} \sim 10^{-48}$ for $\Lambda_\unp = 1~$TeV and $M_\unp = \Mpl$, and alternatively to $\lambda_{\rm min} = 10^{-9}$ when $\Lambda_\unp = \Mgut$ and $M_\unp = \Mpl$.  In both cases when $\du = 3$, the minimum possible black hole mass is on the order of 1~kg

The absolute smallest black hole mass can be found at the intersection of the diagonal bounds and the lower limit $\lambda_{\rm min}(\Lambda_\unp, M_\unp,\du,\dbz)$ hierarchy (Figures~\ref{fig2} and \ref{fig3}).  In the limiting case $\du = \dbz = 3$, one coincidently finds $\Mbh^{\rm min} \sim 1~$kg for all scale energies.    Alternatively, if $\du=3$ and $\dbz=4$, the bounds diverge and $\Mbh^{\rm min} \sim 10^{-8}~$kg for $\Lambda_\unp = 1~$TeV, while $\Mbh^{\rm min} \sim 0.1~$kg for $\Lambda_\unp \sim \Mgut$.  Microscopic black holes ($\Mbh \sim 1~$TeV) would require $\lambda \sim 10^{-56}$ for $\Lambda_\unp = M_{\rm GUT}$, and $\lambda \sim 10^{-81}$ for $\Lambda_\unp = 1~$TeV.  These are obviously not realizable in the vector unparticle framework.

\section{Conclusions}
Vector unparticle fields coupling to the standard model induce repulsive corrections to the gravitational potential that can, in a general relativistic approach, yield multiple Reissner-Nordstr\"om-like horizons without the requirement of electric charge.  Insisting these solutions respect the cosmic censorship hypothesis directly leads to constraints on the vector unparticle coupling constant $\lambda$.   These and more fundamental limits on $\lambda$ from the mass hierarchy between $\Lambda_\unp$ and $M_\unp$, and the scaling dimensions $\du,\dbz$, can be used to theoretically constrain the mass of the smallest possible black hole that can form in a universe governed by this theory.
 
Future potential investigations in this area are ripe.   Unparticle-enhanced to the Kerr metric offer intriguing possibilities of opening up geodesic paths to new universes, without the necessity of electric charge.  This would equivalently introduce new constraints on the possible values of angular momenta that the black holes could assume.   The possibility of naked singularity formation in non-spherical collapse ({\it e.g.} through Thorne's Hoop Conjecture \cite{hoop}) has been investigated and shown to be theoretically possible \cite{shapiro}.   While the metric (\ref{vecmetric}) is inherently spherical, an investigation of unparticle perturbations to such violations of the cosmic censorship hypothesis would be of interest.   

\vskip 1cm
\noindent{\bf Acknowledgments}\\
Thanks to Euro Spallicci for insightful discussions, as well as to Marc Kamionkowski and the generous hospitality of the Moore Center for Theoretical Cosmology and Physics, at which a portion of this research was conducted.  This work was supported in part by the Research Corporation for Science Advancement.

\pagebreak

 \end{document}